# Tunable Perfect Anomalous Reflection in Metasurfaces with Capacitive Lumped Elements

O. Tsilipakos[1], F. Liu[2], A. Pitilakis[1,3], A.C. Tasolamprou[1], D.-H. Kwon[4], M.S. Mirmoosa[2],
N.V. Kantartzis[1,3], E.N. Economou[1], M. Kafesaki[1,5], C.M. Soukoulis[1,6], S.A. Tretyakov[2]

[1] Foundation for Research and Technology Hellas, 71110, Heraklion, Crete, Greece
[2] Department of Electronics and Nanoengineering, Aalto University, P.O. Box 15500, Espoo, Finland
[3] Dept. of Electrical and Computer Engineering, Aristotle University of Thessaloniki, Thessaloniki, Greece
[4] Dept. of Electrical and Computer Engineering, Univ. of Massachusetts Amherst, Amherst, MA 01002, USA
[5] Department of Materials Science and Technology, University of Crete, 71003, Heraklion, Crete, Greece
[6] Ames Laboratory and Department of Physics and Astronomy, Iowa State University, Ames, Iowa 50011, USA
otsilipakos@iesl.forth.gr

*Abstract* – **We demonstrate tunable perfect anomalous reflection with metasurfaces incorporating lumped elements. The tunable capacitance of each element provides continuous control over the local surface reactance, allowing for controlling the evanescent field distribution and efficiently tilting the reflected wavefront away from the specular direction. The performance of the metasurface is evaluated for both TE and TM polarization and for reflection to the first and second diffraction order.**

## I. Introduction

Metasurfaces are ultrathin periodic structures with subwavelength unit cells made of purposefully designed materials and geometries that control the overall electromagnetic (EM) properties and hence their response to impinging fields [1,2]. Precise control over the constituent meta-atoms has enabled the realization of structures with unique and often unnatural properties and has opened the path for exciting applications such as wavefront shaping. One of the most prominent examples is anomalous reflection, where the reflected wave front is steered away from the specular direction. This can be achieved by controlling the local phase imparted by the subwavelength inclusions on the incident field. For example, imposing a linear phase gradient tilts the wavefront according to the generalized Snell's law [3]. However, this approach results in parasitic reflections to undesired directions. Recently, it has been shown with a patch array metasurface that through a strong nonlocal response parasitic reflections can be suppressed and perfect anomalous reflection can be achieved [4].

Here we target *tunable* perfect anomalous reflection by incorporating in the unit cell a tunable lumped element providing variable capacitance. We require continuous control of $C$ [5], allowing for complete control over the reactive response of every meta-atom and thus even greater flexibility compared to metasurfaces with two reflection phase states [6]. In order to achieve high reflection efficiency, we start from the prescription of a linear phase profile and subsequently use optimization to fine tune the design. We demonstrate perfect anomalous reflection to the first or second diffraction order and study both TE and TM polarizations. Controlling, in addition, the resistance of the lumped element we can achieve with the same unit cell design perfect absorption for arbitrary incidence angle, operating frequency and both polarizations.

## II. Tunable Anomalous Reflection with Continuously Controlled RC Loads

The metasurface unit cell under study is depicted in Fig. 1(a). It consists of two copper patches interconnected with a tunable $RC$ element, lying on a metal-backed dielectric substrate. The geometric and material parameters are given in the caption of Fig. 1. In order to achieve perfect anomalous reflection, we construct a supercell of $N$ unit cells (the total extent is $D=Nd_x$) with varying capacitances $C_1, C_2, \ldots C_N$ [Fig. 1(b)] while setting $R$=0.01 Ω for all unit cells to minimize the absorption. For normal incidence and $D < 3\lambda$, four diffraction orders (±1, ±2) besides the specular become propagating; one port is assigned to each diffraction order for measuring the corresponding power and the naming convention is shown in Fig. 1(b). The reflection angle for a given diffraction order, $m$, is given by $k_0(\sin\theta_r - \sin\theta_i) = m(2\pi/D)$ according to momentum conservation. By imposing a linear phase profile along the supercell $\varphi(x) = \varphi_0 + m(2\pi/D)x$ (or $\varphi(y)$ for steering in the $yz$ plane) we can promote a single diffraction order over the remaining leakage channels. In order to specify the required capacitances, we need a "look-up table" relating the reflection phase with the capacitance. We construct it by illuminating the uniform metasurface with a normally-incident plane wave at an operating frequency of 5 GHz [Fig. 1(c)]. To be realistic we have limited the achievable series capacitances in the range [1, 5] pF. Even under this restriction, we access a large reflection phase span of 300 degrees while the reflection amplitude remains very close to unity (inset).

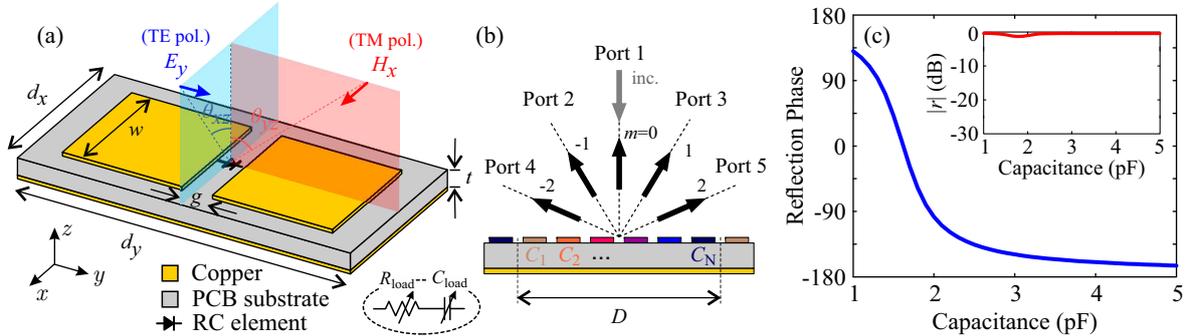

Fig. 1. (a) Metasurface unit cell: Pair of square copper patches on a metal-backed dielectric substrate ($\varepsilon_r$=2.2, tan$\delta$=9·10$^{-4}$) interconnected with a tunable *RC* element. The dimensions are $d_x$=$d_y$/2=9.12 mm, $w$=8.12 mm, $g$=1 mm, and $t$=1.016 mm. The metallization thickness is 17.5 μm. (b) Supercell for anomalous reflection: Port naming convention and correspondence with reflected diffraction orders for normal incidence. (c) Look-up table: Reflection phase for a uniform metasurface as a function of the capacitance of the tunable lumped element. The operating frequency is 5 GHz and the incident wave impinges at normal incidence. The corresponding reflection amplitude is included in the inset.

*A. Incidence in xz plane – TE polarization*

We first focus on incidence in the *xz* plane and TE polarization (**E**=$E_y$**y**), as shown in Fig. 1 (a). We consider a supercell consisting of *N*=8 unit cells stacked along the *x*-axis ($D$=72.96 mm~1.2$\lambda_0$); for normal incidence we get $m = \pm 1$ diffraction orders in the ±55.3º directions. The supercell is effectively described by a three-port network according to Fig. 1 (b). Using the look-up table of Fig. 1 (c), we specify the capacitance of the lumped load for each of the eight unit cells. Subsequently, we extract the S-parameters corresponding to the amplitude of the reflected waves in each of the three ports. Incident power is indeed reflected to the desired diffraction order (port 3), but there is some unwanted radiation in port 2 ($m = -1$) as well. We thus use optimization to further approach perfect anomalous reflection. Specifically, we seek the capacitances that maximize S$_{31}$ (>-0.4 dB) while at the same time minimizing both S$_{21}$ and S$_{11}$ (<-20 dB). The values before and after optimization are:

| $C$ (pF) | $C_1$ | $C_2$ | $C_3$ | $C_4$ | $C_5$ | $C_6$ | $C_7$ | $C_8$ |
|---|---|---|---|---|---|---|---|---|
| Initial | 1.0000 | 1.3362 | 1.5045 | 1.6326 | 1.7709 | 1.9822 | 2.6006 | 5.0000 |
| Optimized | 1.0790 | 1.4078 | 1.5700 | 1.7000 | 1.7001 | 1.9947 | 3.8000 | 5.0000 |

The optimized supercell suppresses radiation towards port 2 and leads to an efficiency of 98% (defined through $|S_{31}|^2$/sum($|S_{x1}|^2$), $x$=1…3) in steering a normally impinging wave towards the +55.3º direction, i.e., port 3 in Fig. 1(b). The full set of S-parameter amplitudes before and after optimization are:

$$|S| = \begin{bmatrix} 0.0781 & 0.2355 & 0.9245 \\ 0.2355 & 0.7961 & 0.3147 \\ 0.9245 & 0.3147 & 0.0128 \end{bmatrix} \xrightarrow{optimization} \begin{bmatrix} 0.0980 & 0.0944 & 0.9470 \\ 0.0944 & 0.8679 & 0.1565 \\ 0.9470 & 0.1565 & 0.1204 \end{bmatrix}, \quad (1)$$

Although the optimization goal was set for normal incidence (S$_{31}$>-0.4 dB), we also get improved performance for incidence from ports 2 and 3. This is also verified by the field plots in Fig. 2, which shows the scattered wavefronts for excitation from all three ports.

The same methodology can be applied for promoting beam deflection to higher diffraction orders. For instance, an *N*=17 supercell ($D$=155 mm~2.6$\lambda_0$) can be used for steering a normally impinging wave towards ±22.8º (first order, ports 3 and 2) as well as ±50.7º (second order, ports 5 and 4). In the latter case, we start from capacitances that correspond to a linear phase profile spanning a 0-4π range (instead of 0-2π range for first-order anomalous reflection) along the supercell. After optimization we achieve a 96% efficiency in steering a normally impinging plane wave to the +50.7º direction (port 5): $|S_{x1}|$=[0.0917, 0.0836, 0.0806, 0.0990, 0.9425], $x$=1…5.

*B. Incidence in yz plane – TM polarization*

We next examine incidence in the $yz$ plane and TM polarization ($\mathbf{H}=H_x\mathbf{x}$) [Fig. 1(a)]. We consider the case of $N=5$ unit cells stacked along the $y$-axis ($D=91.2$ mm$\sim 1.5\lambda_0$) giving rise to $\pm 1$ diffraction orders towards $\pm 41.1°$.

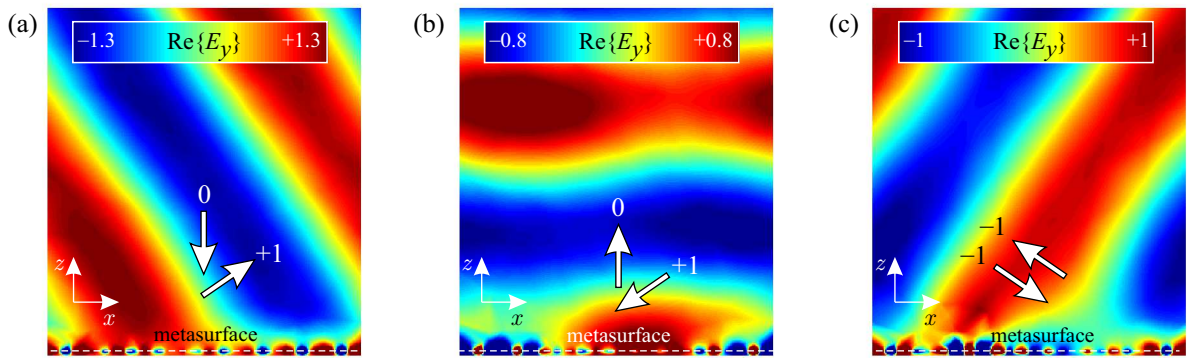

Fig. 2. Anomalous reflection for TE polarization with a metasurface of $N=8$ unit cells along the $x$-axis. The capacitances are optimized. Real part of the scattered electric field for (a) normal incidence (anomalous reflection from port 1 to port 3), (b) incidence from 55.32° (reciprocal to (a)), and (c) incidence from port 2 giving retro-reflected anomalous reflection.

The S-parameter amplitudes before and after optimization are

$$|S|=\begin{bmatrix} 0.1637 & 0.3475 & 0.8892 \\ 0.3475 & 0.7934 & 0.4121 \\ 0.8892 & 0.4121 & 0.0185 \end{bmatrix} \xrightarrow{optimization} \begin{bmatrix} 0.0414 & 0.0532 & 0.9638 \\ 0.0532 & 0.9279 & 0.0963 \\ 0.9638 & 0.0963 & 0.0519 \end{bmatrix}, \qquad (2)$$

The corresponding capacitances are [1.0399, 1.4561, 1.6681, 1.9494, 4.1861] pF and [1.0016, 1.6745, 1.5373, 2.1722, 3.5616] pF, respectively. The efficiencies for excitation from all three ports are exceptionally high: 99.5%, 98.6%, and 98.7%, highlighting the robustness of the proposed design for the TM polarization as well.

## IV. Discussion and Conclusion

We have numerically demonstrated tunable perfect anomalous reflection in metasurfaces that incorporate lumped loads for continuous control over the meta-atoms. Starting from the prescription of a linear phase gradient and subsequently using optimization, we have achieved high efficiencies for anomalous reflection towards the first and second diffraction order for both TE and TM polarizations. The same unit cell design can be used for achieving tunable perfect absorption for impinging waves of different polarizations, incidence angles and frequencies by controlling the capacitance *and* resistance of the lumped elements (in contrast to anomalous reflection, perfect absorption can be realized using a uniform configuration of all the unit cells). For example, at 5 GHz for TE polarization and normal incidence the required values are $C=1.6965$ pF and $R=3.4406$ Ω, respectively. In all cases considered here the required $RC$ values are realistic and can be supplied by voltage-controlled varactor and varistor elements. The tuning mechanism can be easily realized with software, thus paving the way for software-driven intelligent metasurfaces supporting reconfigurable functionalities [5].

## Acknowledgement

This work was supported by the European Union's Horizon 2020 Future Emerging Technologies call (FETOPEN-RIA) under grant agreement no. 736876 (project VISORSURF).